\documentclass[10pt,twocolumn,english,aps,prd,nofootinbib,preprintnumbers]{revtex4}
\usepackage{natbib}
\usepackage{amssymb,amsmath,amsfonts,amsbsy,epsfig,graphicx}
\usepackage[dvipsnames]{xcolor}
\usepackage{multirow}
\usepackage{lipsum}
\usepackage{blindtext}
\usepackage{xspace}
\usepackage{afterpage}
\usepackage{gensymb}
\usepackage{booktabs}

\definecolor{lgray}{gray}{0.35}
\usepackage[colorlinks=true,urlcolor=Gray,linkcolor=lgray,citecolor=Gray, pdfpagelabels=true,hypertexnames=true,plainpages=false,naturalnames=false]{hyperref}
\newcommand{\be}{\begin{equation}}
\newcommand{\ee}{\end{equation}}
\newcommand{\bea}{\begin{eqnarray}}
\newcommand{\eea}{\end{eqnarray}}

\newcommand{\Ha}{H$\alpha$\xspace}
\newcommand{\OII}{$\left[\mathrm{O\,\textrm{\textsc{ii}}}\right]$\xspace}

\def\L{{\mathcal L}}

\begin{document}

\title{Cosmological constraints from galaxy multi-tracers in the nearby Universe}

\author{Ginevra Favole}\email{ginevra.favole@port.ac.uk}
\affiliation{Institute of Cosmology \& Gravitation, University of Portsmouth, Dennis Sciama Building, Portsmouth, PO13FX, UK}

\author{Domenico Sapone}\email{domenico.sapone@uchile.cl}
\author{Javier Silva Lafaurie}\email{javier.silva@ug.uchile.cl}
\affiliation{Grupo de Cosmolog\'ia y Astrof\'isica Te\'orica, Departamento de F\'{i}sica, FCFM, \mbox{Universidad de Chile}, Blanco Encalada 2008, Santiago, Chile}

\begin{abstract}
\noindent The Baryon Acoustic Oscillation (BAO) scale in the clustering of galaxies is a powerful standard ruler to measure cosmological distances and determine the geometry of the Universe. Past surveys have detected the BAO feature in the clustering of different galaxy samples, most of them composed of redder, quiescent galaxies and bluer, star-forming ones out to redshift $z\sim1$. Besides these targets, new upcoming surveys will observe high-redshift galaxies with bright nebular emission lines out to $z\sim2$, quasars and Lyman-$\alpha$ quasars at $z>2$. All these different galaxy targets will be used as multi-tracers of the same underlying dark matter field. By combining them over wide cosmological volumes, we will be able to beat cosmic variance and measure the growth of structure with unprecedented accuracy. 
In this work, we measure the BAO scale in the two-point auto- and cross-correlation functions of three independent populations of multi-tracers extracted from the SDSS DR7 Main galaxy sample at redshift $0.02<z<0.22$. Combining their covariances, we find accurate constraints on the shift parameter $\alpha = 1.00\pm 0.04$ and $D_{\rm V}(z=0.1)/r_{\rm s}=2.92\pm 0.12$.
\end{abstract}

\maketitle

\section{Introduction}
\label{sec:intro}\vspace{-0.2cm}
During the last decades, observations have led to the general acceptance that the Universe is in a phase of accelerated expansion. 
In a homogeneous and isotropic Universe, the simplest way to account for such expansion is to introduce a constant term in the Einstein equations, dubbed as cosmological constant ($\Lambda$). 
Based on recent observations \cite{Aghanim:2018eyx, Eisenstein:2005su, Blake:2011wn} and on the simplicity of the model, the cosmological constant is still the most accepted dark energy candidate responsible for the acceleration of the Universe. 
Despite of its simplicity, such a scenario has raised several theoretical issues, which have led cosmologists to invoke more sophisticated dark energy models without succeeding on the task, see~\cite{Sapone:2010iz}. An alternative approach suggests that we would need to modify the laws of gravity and make it weaker at larger scales to mimic, phenomenologically, the observed expansion~\cite{Kunz:2006ca,Tsujikawa:2010zza}. 

The fundamental observables that trace the dynamics of the Universe are the Hubble parameter $H(z)$ and the angular diameter distance $D_{\rm A}(z)$, which are directly connected to the properties of matter and quantify the overall expansion of the Universe.  
Observational exploration is necessary to provide an indication about the dynamics of the Universe. One way of understanding this is to measure distances at different epochs. Modern cosmology has been revolutionised when the definition of {\em standard ruler} \cite{Seo:2003pu} was introduced: a distance scale in the Universe whose size and evolution with redshift are known. An ideal candidate is the Baryon Acoustic Oscillation scale (BAO; \cite{Eisenstein:2005su}) observed at the last scattering surface in the Cosmic Microwave Background (CMB) radiation. This feature represents the width of the primordial density fluctuations that propagate as acoustic waves in the early baryon-photon fluid. Such a distance can be decomposed into a radial, $H(z)$, and a transverse, $D_A(z)$, direction, which allow us to measure the expansion history of the UniverseGreen:2012mj. 
If the standard cosmological model, i.e. the structures that we see today, have been generated by gravitational collapse of the primordial seeds in an expanding, homogeneous and isotropic Universe, then we should see an excess of baryonic matter in the distribution of galaxies at the same comoving scale. This excess of baryonic matter is visible as a prominent peak around $110\,h^{-1}$Mpc in the galaxy two-point correlation function. The first detection of the BAO peak happened in SDSS~\cite{Eisenstein:2005su} and was then confirmed by 2dFGRS~\cite{Cole:2005sx}, BOSS~\cite{Dawson:2012va} and WiggleZ Dark Energy Survey~\cite{Blake:2011wn}, VIPERS~\cite{delaTorre:2013rpa} and eBOSS~\cite{Dawson:2015wdb}. 

New upcoming surveys, such as the Dark Energy Spectroscopic Instrument (DESI) \cite{Aghamousa:2016zmz, Aghamousa:2016sne}, Euclid \cite{Laureijs:2011gra, Amendola:2016saw}, Subaru Prime Focus Spectrograph (PFS) \cite{Sugai:2012hd, Smee:2012wd}, the Large Synoptic Survey Telescope (LSST) \cite{Abell:2009aa}, or the Wide Field Infrared Survey Telescope (WFIRST) \cite{Green:2012mj, Spergel:2015sza}, will observe tens of hundreds of millions of galaxies positions and spectra covering enormous cosmological volumes and extend the observations at very high redshifts ($z\sim 2-3$). These observations will map the late time dynamics of the Universe with unprecedented precision (few percents on the final cosmological parameters~\cite{Blanchard:2019oqi}). It is therefore imperative to gain as much information as possible from these data sets. 

One statistical limitation of measuring the cosmological parameters is due to the cosmic variance effects in the survey volume. Recently, it was shown  \cite{Abramo:2013awa, Montero-Dorta:2019kyb} that by cross-correlating different dark matter tracers over wide cosmological volumes it is possible to beat cosmic variance, dramatically reducing the uncertainties on the observables. In fact, while the effective volume still remains a limitation, the relative information between different species is not. 
In this paper, we show that using a multi-tracer approach we can lower the errors on the scale distortion parameter $\alpha$. In particular, we use luminous red galaxies and emission line galaxies as tracers to map the dynamics of the Universe at $z\sim 0.1$.

The paper is organised as follow: in Section \ref{sec:data}, we describe the observational samples of multi-tracers used in the analysis; in Section \ref{sec:measurements}, we infer the measurements and observables considered; in Section \ref{sec:methodology}, we explain the methodology used in the analysis. We conclude by presenting and discussing our main results in Section \ref{sec:results}. 

\section{DATA}
\label{sec:data}\vspace{-0.2cm}
We analyse three independent galaxy populations selected from the SDSS DR7 Main galaxy sample \citep[]{Strauss:2002dj}, each one composed of a different tracer, and all of them covering the redshift range $0.02<z<0.22$. Specifically, these samples are: two selections of emission line galaxies (ELGs), one of \OII \citep{Favole_2017} and another one of \Ha  emitters \cite{Favole2019}, plus a selection of luminous red galaxies (LRGs) \cite{Eisenstein:2001cq}. These are the only galaxy multi-tracers currently available in the nearby Universe. The SDSS Main parent sample, which is brighter than $r=17.77$, and covers and effective area of 7300\,deg$^2$ \citep{Guo:2015dda}, was extracted from the NYU-Value Added Galaxy Catalogue\footnote{\url{http://cosmo.nyu.edu/blanton/vagc/}} \citep[]{Blanton:2004aa} and it was spectroscopically matched (i.e. matching the redshifts) to the MPA-JHU\footnote{\url{http://www.mppg.de/SDSS/DR7/}} DR7 release of spectral measurements to assign emission line properties.

We consider only ELGs with well measured spectra, i.e. those with the flag \texttt{ZWARNING\,=\,0}, and with good signal-to-noise, i.e. $\rm{S/N>5}$. Both \OII and \Ha ELG samples have specific star formation rate of $\rm{log(sSFR/M_{\odot}yr^{-1})>-11}$ and line equivalent width of $\rm{EW>10}$\,$\mathring {\rm A}$ to guarantee that we are selecting only very star-forming galaxies. In addition, they are both limited in flux at $2\times10^{-16}$erg\,cm$^{-2}$\,s$^{-1}$ to match the Euclid nominal expected depth and flux limit \citep{Merson:2017efv} at higher redshift. The LRG sample includes only galaxies with $\rm{log(sSFR/M_{\odot}yr^{-1})<-11}$, which are all quiescent. 

The observed (i.e. attenuated by dust) \OII and \Ha ELG luminosities are computed from the corresponding flux densities $F$ as:
\begin{equation}
L\,[{\rm{erg\,\,s^{-1}}}]=4\pi D_{\rm L}^2 10^{-0.4(r_{\rm{p}}-r_{\rm{fib}})}F\,,
\label{eq:lum}
\end{equation}
where $D_{\rm L}$ is the luminosity distance as a function of cosmology and the exponent is the SDSS fibre aperture correction written in terms of the $r$-band petrosian and fibre magnitudes. For further details on the luminosity calculations, we refer the reader to \citep{Favole_2017, Favole2019}.

\begin{figure}
\begin{center}\vspace{-0.3cm}
\includegraphics[width=\linewidth]{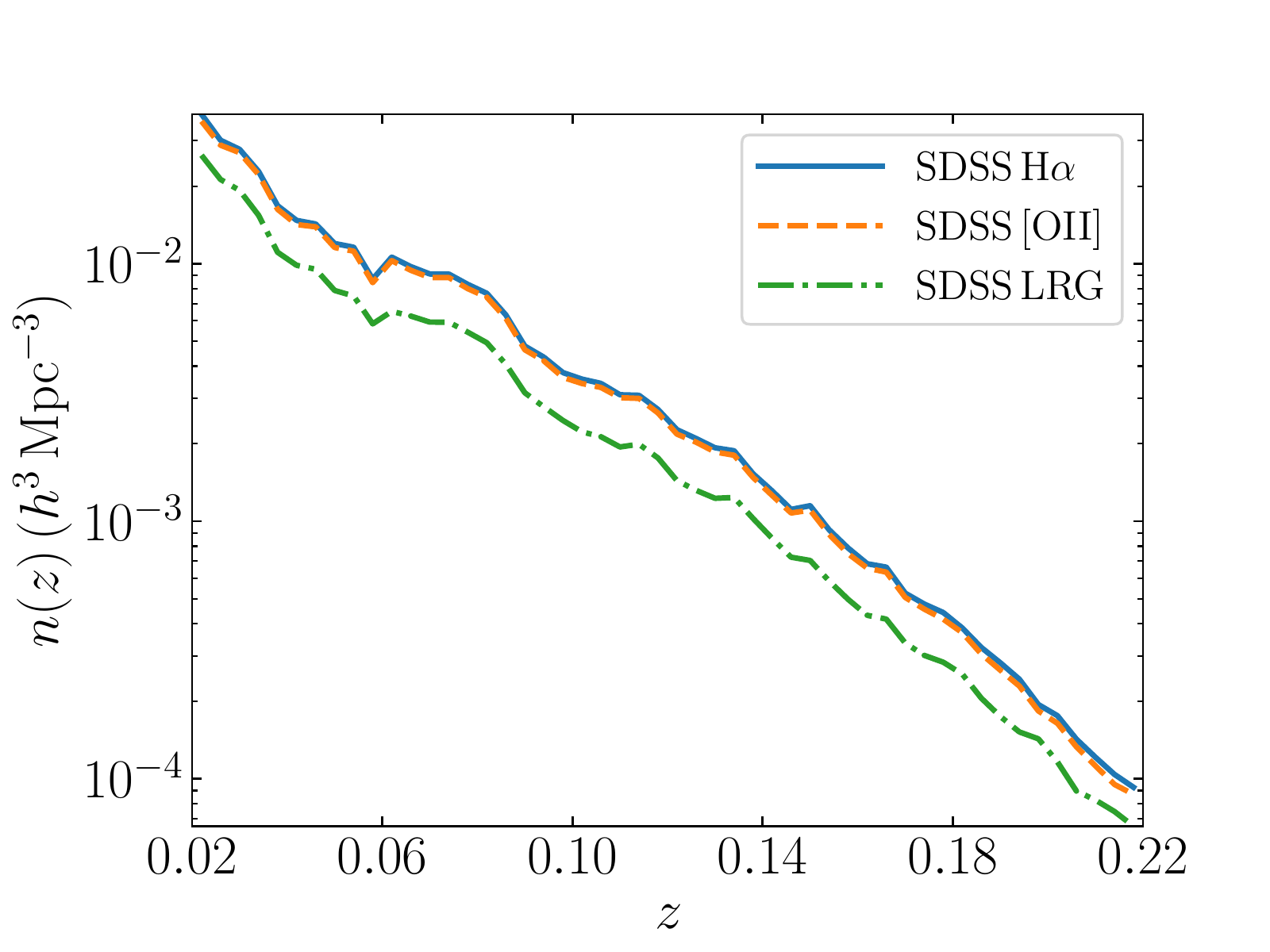}\vspace{-0.3cm}
\caption{Galaxy number density, as a function of redshift, of the SDSS \Ha, \OII and LRG samples at $0.02<z<0.22$.}
\label{fig:nzplot}
\end{center}
\end{figure}

Fig.\,\ref{fig:nzplot} shows the galaxy number density of the SDSS \Ha, \OII and LRG samples as a function of redshift. Compared to the SDSS Main sample in \cite{Ross:2014qpa}, our galaxy selections cover a larger area (we consider both North and South Galactic Caps) and span a slightly different redshift range. Therefore, it is not surprising that our galaxy number densities differ from \cite{Ross:2014qpa}. In particular, our $n(z)$ values are large enough to ensure that the SDSS \Ha, \OII and LRG samples are limited by cosmic variance (i.e., $n(z)\,P(k) > 1$) at $z<0.22$ for $k < 0.095,\,0.093,\,0.071\, h$\,Mpc$^{-1}$, respectively.

\section{Measurements}
\label{sec:measurements}\vspace{-0.2cm}
In Large Scale Structure (LSS) analysis, galaxies can be thought as point-like objects in space-time that move along with the expansion of the Universe, forming bounded structures due to their gravitational interaction. 
By analysing their positions, we can gain information on the underlying theory of gravity. One method consists in quantifying how many objects are present in a given cosmological volume. This method relies on the two-point correlation function (2PCF), which is the excess probability over randoms of finding two galaxies separated by a distance $s$ in redshift-space. 

We measure the two-point auto- and cross-correlation functions of the three galaxy tracers defined in Sec.\,\ref{sec:data} using the Landy and Szalay estimator:
\begin{equation}
    \xi_{\mu \nu}(s)=\frac{D_\mu D_\nu (s) -D_\mu R_\nu (s) -D_\nu R_\mu (s)}{R_\mu R_\nu (s)}+1\,,
\end{equation}
where $s=\sqrt{\pi^2+r_{\rm p}^2}$ represents the redshift-space distance as a function of the parallel ($\pi$) and perpendicular ($r_{\rm p}$) components to the line of sight, while $\mu$ and $\nu$ are the tracers. The DD, DR and RR terms are the normalised and weighted data-data, data-random and random-random pair counts formed from the observed galaxies and the synthetic randoms. We use the equal surface density randoms from the NYU-VAGC. The weighting scheme adopted for the pair counts is
$w=w_{\rm{fc}}w_{\rm{ang}}w_{\rm{FKP}}$ 
for data and $w=w_{\rm{FKP}}$ for randoms. The $w_{\rm{fc}}$ term represents the fibre collision weight (in SDSS fibres cannot be placed closer than 55"). The angular weight $w_{\rm{ang}}=1/f_{\rm{got}}$ accounts for the angular sector completeness, and the FKP \citep[]{Feldman:1993ky} one, 
\begin{equation}
    w_{\rm{FKP}}=\frac{1}{1+\bar{n}(z)P_0}\,,
    \label{eq:fkp}
\end{equation}
corrects for any fluctuation in the number density of tracers. In Eq.\,\eqref{eq:fkp}, $\bar{n}(z)$ is the expected number density of a galaxy at redshift $z$ and we set $P_0=16000\,h^{-3}$Mpc$^3$, which is close to the amplitude of the SDSS power spectrum at $k=0.1h$Mpc$^{-1}$ \citep{Ross:2014qpa}.

We estimate the uncertainties on the SDSS clustering measurements via 200 jackknife re-samplings \citep[][]{miller74, Norberg2009, Norberg:2011ef, Guo:2012nk, Favole:2015rza} 
containing about the same number of data (randoms) each. The covariance matrix for each 2PCF is calculated as \citep{1982jbor.book.....E, Norberg2009, Favole:2015rza}:
\begin{equation}
   \hat{C}_{ij}=\frac{N_{\rm{res}}-1}{N_{\rm{res}}}\sum_{a=1}^{N_{\rm{res}}}(\xi_i^a-\bar{\xi}_i^a)(\xi_j^a-\bar{\xi}_j^a)\,,
   \label{eq:covmatrix}
\end{equation}
where the pre-factor takes into account that in every re-sampling $N_{\rm{res}}-2$ sub-volumes are the same \citep{Norberg:2011ef}, and $\bar{\xi}_i$ is the mean jackknife correlation function in the $i^{\rm{th}}$ bin:
\begin{equation}
\bar{\xi}_i=\sum_{a=1}^{N_{\rm{res}}}\xi_i^a/N_{\rm{res}}\,.    
\end{equation}
The full covariance matrix for all the tracers is built by combining the individual ones in Eq.\,\eqref{eq:covmatrix} as:
\begin{equation}
\label{entire_cov_matrix}
\hat{C} = 
\begin{pmatrix}
    \hat{C}_{\rm{H\alpha-H\alpha}} & \hat{C}_{\rm{H\alpha-[OII]}} & \hat{C}_{\rm{H\alpha-LRG}} \\
    \hat{C}_{\rm{H\alpha-[OII]}} & \hat{C}_{\rm{[OII]-[OII]}}         & \hat{C}_{\rm{[OII]-LRG}} \\
    \hat{C}_{\rm{H\alpha-LRG}}    & \hat{C}_{\rm{[OII]-LRG}}     & \hat{C}_{\rm{LRG-LRG}}        \\
\end{pmatrix}\,.
\end{equation}
The inverse of the covariance matrix, the so-called \textquoteleft\textquoteleft precision matrix" $\hat{\Psi}\equiv\hat{C}^{-1}$, requires some corrections. In fact, Eq.~\eqref{entire_cov_matrix} is obtained from a limited set of re-samplings, $N_{\rm{res}}$, and it has an associated error which propagates into the precision matrix. Following~\cite{Paz:2015kwa}, we implement two corrections to obtain an unbiased estimate of the precision matrix and to reduce the noise in its off-diagonal terms. The bias correction consists in multiplying $\hat{C}^{-1}$ by the Hartlap factor~\cite{Hartlap:2006kj}, which accounts for the limited number of re-samplings and the number of bins $n_{\rm{b}}$ in our measurements of $\xi$: 
\begin{equation}
    \hat{\Psi} = \left(1-\frac{n_{\rm{b}}+1}{N_{\rm{res}}-1}\right)\hat{C}^{-1}\,,
\end{equation}
\begin{figure}
\begin{center}
\includegraphics[width=0.95\linewidth]{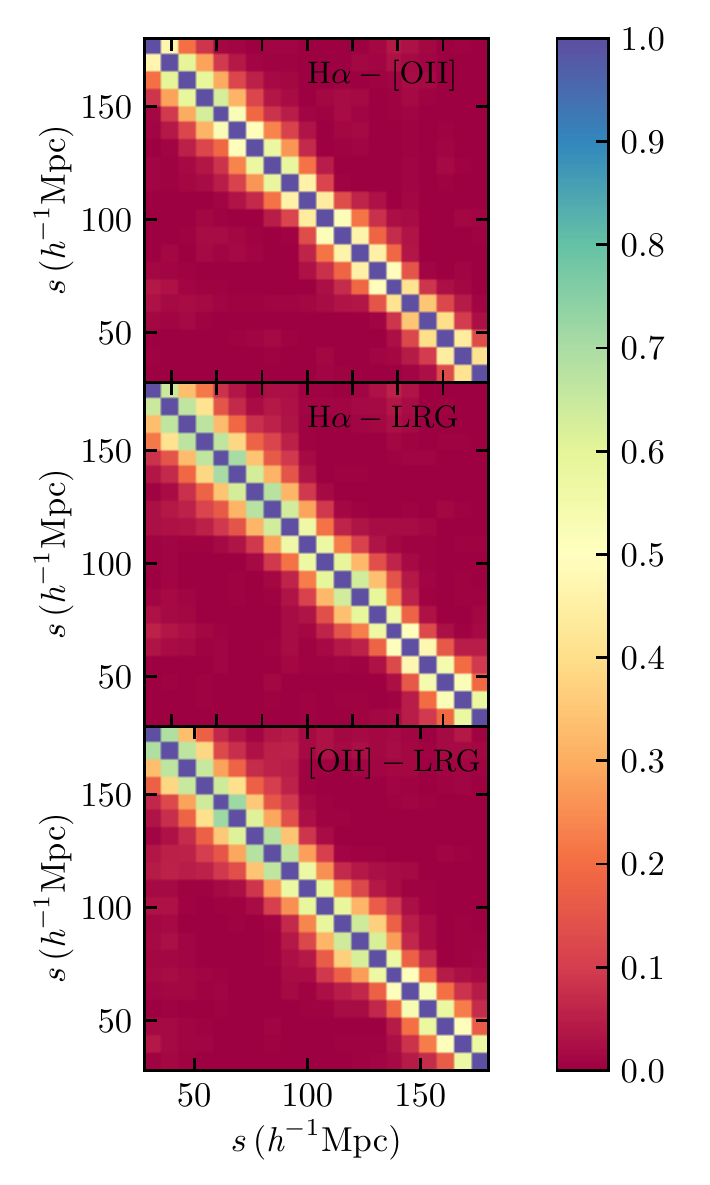}\vspace{-0.5cm}
\caption{Normalised covariance matrices obtained from the monopole cross-correlation functions of the SDSS galaxy tracers. From top to bottom: \Ha-\OII, \Ha-LRG and \OII-LRG.}
\label{fig:monopole}
\end{center}
\end{figure}
The noise correction, also known as \textquoteleft\textquoteleft covariance tapering"~\cite{Kaufman:2008}, can be applied to both covariance and precision matrices. The idea of this method is to neglect the correlation between data pairs far apart through a kernel function of the Mat\'ern  class. Such a correction relies on the tapering matrix $T_{ij}\equiv K\left(||s_i-s_j||\right)$, which is defined as a monoparametric Kernel function \cite{Paz:2015kwa, Wendland1995, WENDLAND1998258} depending on the physical scale of the tracers that we are correlating. This kernel also includes a tapering parameter $T_{\rm{p}}$, which identifies the interval where $K(x)$ takes non-zero values, guaranteeing a vanishing correlation between pairs for larger distances.
The final corrected precision matrix is: 
\begin{equation}
\label{final_pres_matrix}
    \hat{\Psi} = \left(1-\frac{n_{\rm{b}}+1}{N_{\rm{res}}-1}\right)\left(\hat{C}\circ T\right)^{-1}\circ T\,,
\end{equation}
where the $\circ$ symbol indicates the Hadamard product. In our analysis, we assume a tapering parameter $T_{\rm{p}}=50\,h^{-1}$Mpc to ensure that the entire covariance matrix is positive semi-definite. Fig. \ref{fig:monopole} shows the normalised covariance matrices obtained from the monopole cross-correlation functions of the SDSS \Ha-\OII, \Ha-LRG and \OII-LRG multi-tracers.


\section{Methodology}
\label{sec:methodology}\vspace{-0.2cm}
The real-space two-point correlation function $\xi(r)$ is the spatially isotropic Fourier transform of the matter power spectrum $P(k)$ defined as:
\be
\label{eq:Pk_to_xi}
\xi(r)=\frac{1}{2\pi^2} \int P(k) \frac{\sin(kr)}{kr}k^2 {\rm d}k\,.
\ee
The position of the BAO peak inferred from Eq.~\eqref{eq:Pk_to_xi} is expected to appear around $110\,h^{-1}$Mpc, which is well beyond the scales of virialised objects. This implies that the non-linear gravitational effects can be safely ignored. 
For the power spectrum we use the template~\cite{Padmanabhan:2008ag}:
\begin{equation}
\label{eq:Pk_non_lin}   
    P(k)=\left[P_{\rm{lin}}(k)-P_{\rm{dw}}(k)\right]e^{-k^2\Sigma_{\rm{nl}}^2/2}+P_{\rm{dw}}(k)\,,
\end{equation}
where $P_{\rm{lin}}(k)$ is the linear matter power spectrum calculated using the Boltzmann code CLASS~\cite{Lesgourgues:2011re}, and $P_{\rm dw}(k)$ is the de-wiggled power spectrum~\cite{Eisenstein:1997ik}, both using Planck 2015~\cite{Ade:2015xua} fiducial cosmology. The $\Sigma_{\rm nl}$ parameter accounts for the smoothing of the BAO peak due to non-linear effects~\cite{Crocce:2005xy}. 

We compute the theoretical correlation functions needed to fit the SDSS multi-tracer measurements by applying Eq.~\eqref{eq:Pk_to_xi} with the power spectrum given in Eq.~\eqref{eq:Pk_non_lin}. We model the BAO signal as~\cite{Xu:2012hg}: 
\be
\label{fit_model}
\xi_{\rm{model}}(s) = B\xi(\alpha s)+a_0+\frac{a_1}{s}+ \frac{a_2}{s^2}\,,
\ee
where $a_1,a_2,a_3$ are linear nuisance parameters and $B$ accounts for all possible effects on the clustering amplitude, such as the linear bias, the normalisation of the power spectrum, $\sigma_8$, and the redshift space distortions~\cite{Xu:2012hg}. 
In addition, we introduce the shift parameter $\alpha$ which takes into account the distortion between distances measured in the data due to the fiducial cosmology chosen to build the estimator. This is defined as~\cite{Padmanabhan:2008ag}:
\be
\label{shift_parameter}
\alpha = \frac{D_{\rm V}}{r_{\rm s}}\frac{r_s^{\rm{fid}}}{D_V^{\rm{fid}}}\,,
\ee
where $r_s$ represents the sound-horizon~\cite{Hu:1995en, Eisenstein:1997ik} and $D_V$ is the volume-averaged distance defined as  \cite{Eisenstein:2005su}:
\be
\label{volume-averaged-distance}
D_{\rm V}(z)=\left[cz(1+z)^2D_{\rm A}^2(z)H^{-1}(z)\right]^{1/3}, 
\ee
where $D_{\rm A}(z)$ and $H(z)$ are the angular diameter distance and the Hubble parameter at redshift $z$, respectively.

We use a Monte Carlo Markov Chain (MCMC) based on a Metropolis-Hastings algorithm\footnote{https://emcee.readthedocs.io/en/stable/} to find the optimal parameter values. We assume a likelihood function of the form $\L\propto\exp(-\chi^2/2)$, where the $\chi^2$ is computed as:
\be 
\chi^2(\alpha,B) = (\vec{\xi}_{\rm{model}}-\vec{\xi}_{\rm{obs}})^T\hat{\Psi}(\vec{\xi}_{\rm{model}}-\vec{\xi}_{\rm{obs}}).\label{eq:chi2}
\ee
In the equation above, $\vec{\xi}_{\rm{model}}$ is the theoretical correlation function given in Eq.~\eqref{fit_model}, $\vec{\xi}_{\rm{obs}}$ is the observed one, both grouped in a vector at each position, and $\hat{\Psi}$ is the precision matrix given in Eq.~\eqref{final_pres_matrix}.

\section{Results and discussions}
\label{sec:results}\vspace{-0.2cm}
\begin{figure}
\begin{center}
\includegraphics[width=\linewidth]{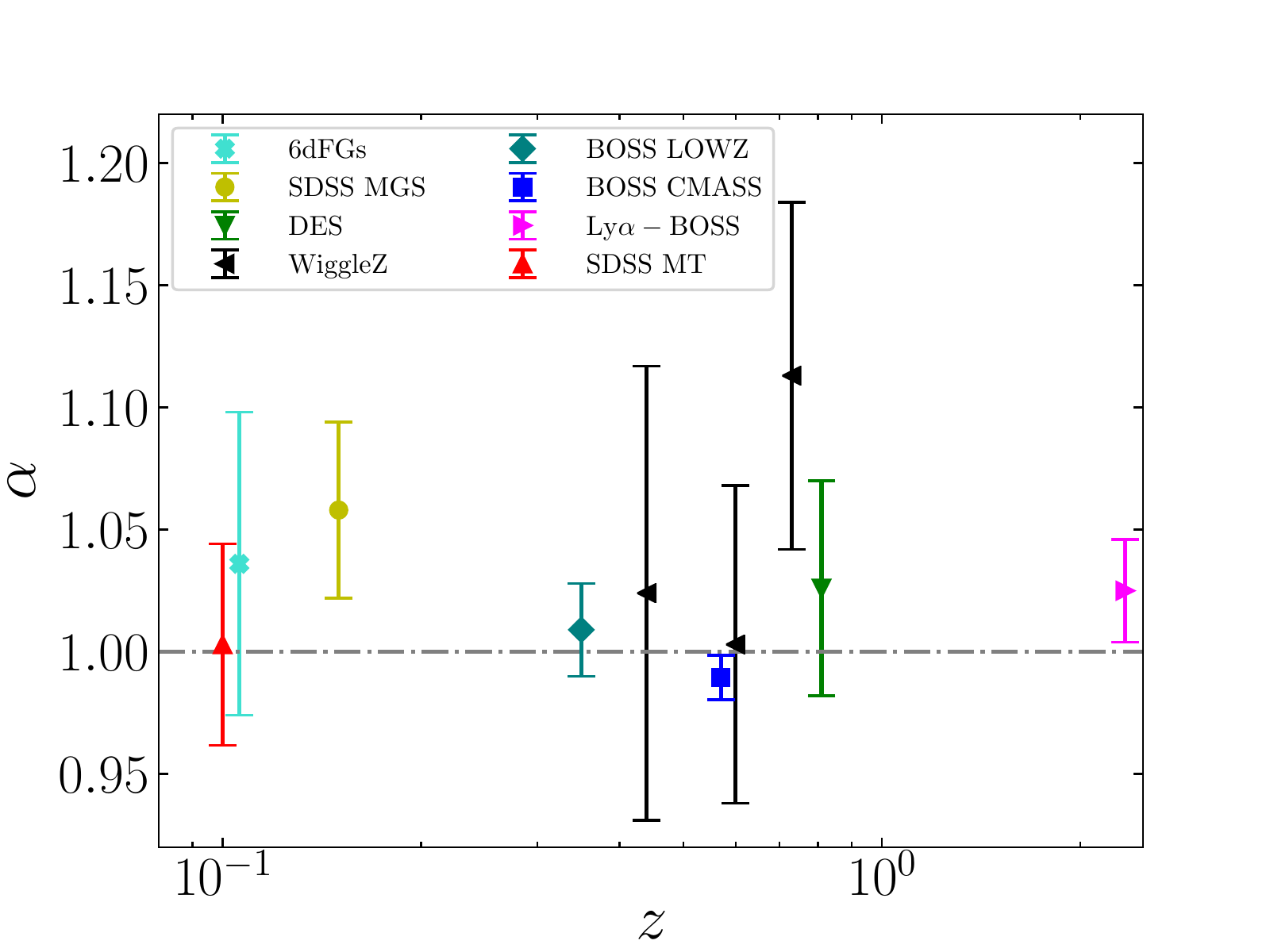}\vspace{-0.3cm}
\caption{Shift parameter $\alpha$ as a function of redshift from different BAO measurements: 6DFGs~\cite{Beutler:2011hx}, MGS~\cite{Ross:2014qpa},  DES~\cite{Abbott:2017wcz}, WiggleZ~\cite{Blake:2011wn}, Lowz-BOSS~\cite{Gil-Marin:2015nqa}, CMASS-BOSS~\cite{Xu:2012hg} and  Ly$\alpha$-BOSS~\cite{Delubac:2014aqe}}.
\label{fig:alpha_plot}
\end{center}
\end{figure}

In this analysis, we have considered different model scenarios with an increasing level of complexity. Our main results are summarised in Tab.~\ref{tab:results_params} and shown in Fig.~\ref{fig:alpha_plot}, together with previous results from literature. 
  First, we use the model given in Eq.~\eqref{fit_model}, which has 5 parameters (5p) common to all the targets: ($B$, $\alpha$, $a_0$, $a_1$, $a_2$). This is equivalent to assume that all the targets respond in the same way to the gravitational interaction and expansion. The second model we test is a modification of Eq.~\eqref{fit_model}, with 13 independent parameters (13p): $\alpha$, common to all the targets, plus three different sets of ($B$, $a_0$, $a_1$, $a_2$). The last model used is again a modification of Eq.~\eqref{fit_model}, with three different sets of ($B$, $\alpha$, $a_0$, $a_1$, $a_2$) i.e., 15 parameters (15p) in total. As a test, we also report the analysis performed using only LRG target. 
In the 5p scenario, we find a shift parameter of $\alpha = 1.00\pm 0.04$, while in the 13p model $\alpha = 1.01 \pm 0.04$.
In the 15p scenario, we find $\alpha = 1.02\pm 0.04$ for both \Ha and \OII, and $\alpha = 0.97\pm 0.05$ for LRGs. The latter is consistent with $\alpha =0.96\pm 0.07$ found using only LRGs, which is directly comparable to the results at $z\sim0.15$ by \cite{Ross:2014qpa} and at $z\sim0.35$ by \cite{chuang2012}.

\begin{table}[t]
\begin{center}
\begin{tabular}{lcccc}
\hline
Models: & 5p & 13p & 15p & LRG \\
\hline
$\alpha_{H_\alpha}$ & - & - &$1.02^{0.04}_{0.03}$& - \\
$\alpha_{OII}$ & - & - &$1.02^{0.04}_{0.04}$ &  -\\
$\alpha_{LRG}$ & - & - &$0.97^{0.05}_{0.05}$ & $0.96^{0.07}_{0.06}$\\
$\alpha$      &$1.00^{0.04}_{0.04}$& $1.01^{0.04}_{0.04}$& - & -\\
\hline
\end{tabular}
\end{center}
\caption{Best-fit constraints from our models.}
\label{tab:results_params}
\end{table}
For our fiducial cosmology, we find a volume-averaged distance of $D_{\rm V}^{\rm{fid}}(z=0.1)=429.90\,$Mpc, and $D_{\rm V}^{\rm{fid}}(z=0.1)/r_{\rm s}^{\rm{fid}}=2.92$. By combining Eq.~\eqref{shift_parameter} with the constraints obtained on $\alpha$, we find that the 1$\sigma$ uncertainty on $D_{\rm V}/r_{\rm s}$ is $\sim0.12$. Assuming $r_{\rm s}=147.41\,$Mpc \cite{Ade:2015xua}, we find $D_{\rm V} = (435.07\pm 17.14)\,$Mpc.

Our study relies on the jackknife covariance matrices from SDSS data, corrected from bias and noise (see Sec.~\ref{sec:measurements}), whereas \cite{Ross:2014qpa} use covariances from synthetic mock catalogues. As shown by \cite{Norberg2009} and \cite{Philcox:2019ued}, jackknife returns reliable covariance estimates only on large scales, that are the scales of interest in our cosmological analysis. Hence, we do not expect our results to change substantially if covariance matrices from mocks were used.  

Another difference between our analysis and \cite{Ross:2014qpa} is the fact that we do not reconstruct the density field. The main idea of BAO reconstruction \citep{Eisenstein:2006nk, Padmanabhan:2012hf} is to smooth the linear matter density field and to sharpen the acoustic peak in the correlation function. This method has the advantage of accurately constraining the non-linear parameter $\Sigma_{\rm{nl}}$. In our analysis we find $\Sigma_{\rm{nl}} \sim 20\,{\rm Mpc}\,h^{-1}$, while \cite{Ross:2014qpa} find $\Sigma_{\rm{nl}}\sim 5\,{\rm Mpc}\,h^{-1}$. A lower value of $\Sigma_{\rm{nl}}$ provides a better signal and tighter constraints on both $B$ and $\alpha$. Our results are fairly compared and in agreement within $1\sigma$ with the pre-reconstruction value of $\alpha = 1.01\pm 0.09$ from \cite{Ross:2014qpa} for LRGs only. The uncertainty we find on $\alpha$ using our multi-tracer analysis is 0.04, identical to the post-reconstruction estimate by \cite{Ross:2014qpa}, and 44\% smaller than their pre-reconstruction value. Hence, we expect that by implementing reconstruction on mocks for galaxy multi-tracers, we will be able to significantly reduce our current error. This result highlights the great potentiality of combining different tracers to constrain more accurately the cosmological parameters.

We remind the reader that for all the models used in this work we assumed flat priors, differently from \cite{Ross:2014qpa}, where Gaussian priors are considered. Flat priors are less informative, but they do not rule out any region of the parameter space.

We have performed a cosmological analysis on the LRG, \Ha and \OII ELG multi-tracers currently available at $z\sim0.1$. For the future, we plan to extend the multi-tracer methodology tested here to the upcoming data sets from the new spectroscopic surveys, such as DESI or Euclid. Our ultimate goal is to include synthetic mock catalogues for galaxy multi-tracers testing the impact of BAO reconstruction on our results. This will enable us to improve the multi-tracer covariance estimates on all scales, and hopefully we will be able to put tight constraints on the non-linear redshift-space distortions.


\section{Acknowledgments}
\label{sec:thanks}
\vspace{-0.2cm}
GF is funded through a Dennis Sciama fellowship at the Institute of Cosmology and Gravitation (ICG), at Portsmouth University. 
JS acknowledges financial support from CONICYT.
The authors are grateful to Melita Carbone and Philipp Sudek for insightful discussions and to Daniel Eisenstein for providing useful comments during the preparation of this work.

\bibliographystyle{utphys}
\bibliography{biblio_corr}

\providecommand{\href}[2]{#2}\begingroup\raggedright\begin{thebibliography}{10}

\bibitem{Aghanim:2018eyx}
{\bfseries Planck} Collaboration, N.~Aghanim {\em et~al.}, ``{Planck 2018
  results. VI. Cosmological parameters},''
\href{http://arxiv.org/abs/1807.06209}{{\ttfamily arXiv:1807.06209
  [astro-ph.CO]}}.

\bibitem{Eisenstein:2005su}
{\bfseries SDSS} Collaboration, D.~J. Eisenstein {\em et~al.}, ``{Detection of
  the Baryon Acoustic Peak in the Large-Scale Correlation Function of SDSS
  Luminous Red Galaxies},'' \href{http://dx.doi.org/10.1086/466512}{{\em
  Astrophys. J.} {\bfseries 633} (2005) 560--574},
\href{http://arxiv.org/abs/astro-ph/0501171}{{\ttfamily arXiv:astro-ph/0501171
  [astro-ph]}}.

\bibitem{Blake:2011wn}
C.~Blake {\em et~al.}, ``{The WiggleZ Dark Energy Survey: testing the
  cosmological model with baryon acoustic oscillations at z=0.6},''
  \href{http://dx.doi.org/10.1111/j.1365-2966.2011.19077.x}{{\em Mon. Not. Roy.
  Astron. Soc.} {\bfseries 415} (2011) 2892--2909},
\href{http://arxiv.org/abs/1105.2862}{{\ttfamily arXiv:1105.2862
  [astro-ph.CO]}}.

\bibitem{Sapone:2010iz}
D.~Sapone, ``{Dark Energy in Practice},''
  \href{http://dx.doi.org/10.1142/S0217751X10050743}{{\em Int. J. Mod. Phys.}
  {\bfseries A25} (2010) 5253--5331},
\href{http://arxiv.org/abs/1006.5694}{{\ttfamily arXiv:1006.5694
  [astro-ph.CO]}}.

\bibitem{Kunz:2006ca}
M.~Kunz and D.~Sapone, ``{Dark Energy versus Modified Gravity},''
  \href{http://dx.doi.org/10.1103/PhysRevLett.98.121301}{{\em Phys. Rev. Lett.}
  {\bfseries 98} (2007) 121301},
\href{http://arxiv.org/abs/astro-ph/0612452}{{\ttfamily arXiv:astro-ph/0612452
  [astro-ph]}}.

\bibitem{Tsujikawa:2010zza}
S.~Tsujikawa, ``{Modified gravity models of dark energy},''
  \href{http://dx.doi.org/10.1007/978-3-642-10598-2_3}{{\em Lect. Notes Phys.}
  {\bfseries 800} (2010) 99--145},
\href{http://arxiv.org/abs/1101.0191}{{\ttfamily arXiv:1101.0191 [gr-qc]}}.

\bibitem{Seo:2003pu}
H.-J. Seo and D.~J. Eisenstein, ``{Probing dark energy with baryonic acoustic
  oscillations from future large galaxy redshift surveys},''
  \href{http://dx.doi.org/10.1086/379122}{{\em Astrophys. J.} {\bfseries 598}
  (2003) 720--740},
\href{http://arxiv.org/abs/astro-ph/0307460}{{\ttfamily arXiv:astro-ph/0307460
  [astro-ph]}}.

\bibitem{Cole:2005sx}
{\bfseries 2dFGRS} Collaboration, S.~Cole {\em et~al.}, ``{The 2dF Galaxy
  Redshift Survey: Power-spectrum analysis of the final dataset and
  cosmological implications},''
  \href{http://dx.doi.org/10.1111/j.1365-2966.2005.09318.x}{{\em Mon. Not. Roy.
  Astron. Soc.} {\bfseries 362} (2005) 505--534},
\href{http://arxiv.org/abs/astro-ph/0501174}{{\ttfamily arXiv:astro-ph/0501174
  [astro-ph]}}.

\bibitem{Dawson:2012va}
{\bfseries BOSS} Collaboration, K.~S. Dawson {\em et~al.}, ``{The Baryon
  Oscillation Spectroscopic Survey of SDSS-III},''
  \href{http://dx.doi.org/10.1088/0004-6256/145/1/10}{{\em Astron. J.}
  {\bfseries 145} (2013) 10},
\href{http://arxiv.org/abs/1208.0022}{{\ttfamily arXiv:1208.0022
  [astro-ph.CO]}}.

\bibitem{delaTorre:2013rpa}
S.~de~la Torre {\em et~al.}, ``{The VIMOS Public Extragalactic Redshift Survey
  (VIPERS). Galaxy clustering and redshift-space distortions at z=0.8 in the
  first data release},''
  \href{http://dx.doi.org/10.1051/0004-6361/201321463}{{\em Astron. Astrophys.}
  {\bfseries 557} (2013) A54},
\href{http://arxiv.org/abs/1303.2622}{{\ttfamily arXiv:1303.2622
  [astro-ph.CO]}}.

\bibitem{Dawson:2015wdb}
K.~S. Dawson {\em et~al.}, ``{The SDSS-IV extended Baryon Oscillation
  Spectroscopic Survey: Overview and Early Data},''
  \href{http://dx.doi.org/10.3847/0004-6256/151/2/44}{{\em Astron. J.}
  {\bfseries 151} (2016) 44},
\href{http://arxiv.org/abs/1508.04473}{{\ttfamily arXiv:1508.04473
  [astro-ph.CO]}}.

\bibitem{Aghamousa:2016zmz}
{\bfseries DESI} Collaboration, A.~Aghamousa {\em et~al.}, ``{The DESI
  Experiment Part I: Science,Targeting, and Survey Design},''
\href{http://arxiv.org/abs/1611.00036}{{\ttfamily arXiv:1611.00036
  [astro-ph.IM]}}.

\bibitem{Aghamousa:2016sne}
{\bfseries DESI} Collaboration, A.~Aghamousa {\em et~al.}, ``{The DESI
  Experiment Part II: Instrument Design},''
\href{http://arxiv.org/abs/1611.00037}{{\ttfamily arXiv:1611.00037
  [astro-ph.IM]}}.

\bibitem{Laureijs:2011gra}
{\bfseries EUCLID} Collaboration, R.~Laureijs {\em et~al.}, ``{Euclid
  Definition Study Report},''
\href{http://arxiv.org/abs/1110.3193}{{\ttfamily arXiv:1110.3193
  [astro-ph.CO]}}.

\bibitem{Amendola:2016saw}
L.~Amendola {\em et~al.}, ``{Cosmology and fundamental physics with the Euclid
  satellite},'' \href{http://dx.doi.org/10.1007/s41114-017-0010-3}{{\em Living
  Rev. Rel.} {\bfseries 21} no.~1, (2018) 2},
\href{http://arxiv.org/abs/1606.00180}{{\ttfamily arXiv:1606.00180
  [astro-ph.CO]}}.

\bibitem{Sugai:2012hd}
H.~Sugai {\em et~al.}, ``{Prime focus spectrograph: Subaru's future},''
  \href{http://dx.doi.org/10.1117/12.926954}{{\em Proc. SPIE Int. Soc. Opt.
  Eng.} {\bfseries 8446} (2012) 84460Y},
\href{http://arxiv.org/abs/1210.2719}{{\ttfamily arXiv:1210.2719
  [astro-ph.IM]}}.

\bibitem{Smee:2012wd}
S.~Smee {\em et~al.}, ``{The Multi-Object, Fiber-Fed Spectrographs for SDSS and
  the Baryon Oscillation Spectroscopic Survey},''
  \href{http://dx.doi.org/10.1088/0004-6256/146/2/32}{{\em Astron. J.}
  {\bfseries 146} (2013) 32},
\href{http://arxiv.org/abs/1208.2233}{{\ttfamily arXiv:1208.2233
  [astro-ph.IM]}}.

\bibitem{Abell:2009aa}
{\bfseries LSST Science, LSST Project} Collaboration, P.~A. Abell {\em et~al.},
  ``{LSST Science Book, Version 2.0},''
\href{http://arxiv.org/abs/0912.0201}{{\ttfamily arXiv:0912.0201
  [astro-ph.IM]}}.

\bibitem{Green:2012mj}
J.~Green {\em et~al.}, ``{Wide-Field InfraRed Survey Telescope (WFIRST) Final
  Report},''
\href{http://arxiv.org/abs/1208.4012}{{\ttfamily arXiv:1208.4012
  [astro-ph.IM]}}.

\bibitem{Spergel:2015sza}
D.~Spergel {\em et~al.}, ``{Wide-Field InfrarRed Survey Telescope-Astrophysics
  Focused Telescope Assets WFIRST-AFTA 2015 Report},''
\href{http://arxiv.org/abs/1503.03757}{{\ttfamily arXiv:1503.03757
  [astro-ph.IM]}}.

\bibitem{Blanchard:2019oqi}
{\bfseries Euclid} Collaboration, A.~Blanchard {\em et~al.}, ``{Euclid
  preparation: VII. Forecast validation for Euclid cosmological probes},''
\href{http://arxiv.org/abs/1910.09273}{{\ttfamily arXiv:1910.09273
  [astro-ph.CO]}}.

\bibitem{Abramo:2013awa}
L.~R. Abramo and K.~E. Leonard, ``{Why multi-tracer surveys beat cosmic
  variance},'' \href{http://dx.doi.org/10.1093/mnras/stt465}{{\em Mon. Not.
  Roy. Astron. Soc.} {\bfseries 432} (2013) 318},
\href{http://arxiv.org/abs/1302.5444}{{\ttfamily arXiv:1302.5444
  [astro-ph.CO]}}.

\bibitem{Montero-Dorta:2019kyb}
A.~D. Montero-Dorta, L.~R. Abramo, B.~R. Granett, S.~de~la Torre, and L.~Guzzo,
  ``{The Multi-Tracer Optimal Estimator applied to VIPERS},''
\href{http://arxiv.org/abs/1909.00010}{{\ttfamily arXiv:1909.00010
  [astro-ph.CO]}}.

\bibitem{Strauss:2002dj}
{\bfseries SDSS} Collaboration, M.~A. Strauss {\em et~al.}, ``{Spectroscopic
  Target Selection in the Sloan Digital Sky Survey: The Main Galaxy Sample},''
  \href{http://dx.doi.org/10.1086/342343}{{\em Astron. J.} {\bfseries 124}
  (2002) 1810},
\href{http://arxiv.org/abs/astro-ph/0206225}{{\ttfamily arXiv:astro-ph/0206225
  [astro-ph]}}.

\bibitem{Favole_2017}
G.~Favole, S.~A. Rodr\'{i}guez-Torres, J.~Comparat, F.~Prada, H.~Guo,
  A.~Klypin, and A.~D. Montero-Dorta, ``Galaxy clustering dependence on the
  {$\left[\mathrm{O\,\textrm{\textsc{ii}}}\right]$\xspace} emission line
  luminosity in the local universe,''
  \href{http://dx.doi.org/10.1093/mnras/stx1980}{{\em Monthly Notices of the
  Royal Astronomical Society} {\bfseries 472} no.~1, (Aug, 2017) 550--558}.

\bibitem{Favole2019}
G.~Favole, ``{The clustering of H$\alpha$ emitters in the nearby Universe},''
  {\em in preparation} (2019) .

\bibitem{Eisenstein:2001cq}
{\bfseries SDSS} Collaboration, D.~J. Eisenstein {\em et~al.}, ``{Spectroscopic
  target selection for the Sloan Digital Sky Survey: The Luminous red galaxy
  sample},'' \href{http://dx.doi.org/10.1086/323717}{{\em Astron. J.}
  {\bfseries 122} (2001) 2267},
\href{http://arxiv.org/abs/astro-ph/0108153}{{\ttfamily arXiv:astro-ph/0108153
  [astro-ph]}}.

\bibitem{Guo:2015dda}
H.~Guo {\em et~al.}, ``{Redshift-space clustering of SDSS galaxies - luminosity
  dependence, halo occupation distribution, and velocity bias},''
  \href{http://dx.doi.org/10.1093/mnras/stv1966}{{\em Mon. Not. Roy. Astron.
  Soc.} {\bfseries 453} no.~4, (2015) 4368--4383},
\href{http://arxiv.org/abs/1505.07861}{{\ttfamily arXiv:1505.07861
  [astro-ph.CO]}}.

\bibitem{Blanton:2004aa}
{\bfseries SDSS} Collaboration, M.~R. Blanton {\em et~al.}, ``{NYU-VAGC: A
  Galaxy catalog based on new public surveys},''
  \href{http://dx.doi.org/10.1086/429803}{{\em Astron. J.} {\bfseries 129}
  (2005) 2562--2578},
\href{http://arxiv.org/abs/astro-ph/0410166}{{\ttfamily arXiv:astro-ph/0410166
  [astro-ph]}}.

\bibitem{Merson:2017efv}
A.~Merson, Y.~Wang, A.~Benson, A.~Faisst, D.~Masters, A.~Kiessling, and
  J.~Rhodes, ``{Predicting H$\alpha$ emission-line galaxy counts for future
  galaxy redshift surveys},''
  \href{http://dx.doi.org/10.1093/mnras/stx2649}{{\em Mon. Not. Roy. Astron.
  Soc.} {\bfseries 474} no.~1, (2018) 177--196},
\href{http://arxiv.org/abs/1710.00833}{{\ttfamily arXiv:1710.00833
  [astro-ph.GA]}}.

\bibitem{Ross:2014qpa}
A.~J. Ross, L.~Samushia, C.~Howlett, W.~J. Percival, A.~Burden, and M.~Manera,
  ``{The clustering of the SDSS DR7 main Galaxy sample ? I. A 4 per cent
  distance measure at $z = 0.15$},''
  \href{http://dx.doi.org/10.1093/mnras/stv154}{{\em Mon. Not. Roy. Astron.
  Soc.} {\bfseries 449} no.~1, (2015) 835--847},
\href{http://arxiv.org/abs/1409.3242}{{\ttfamily arXiv:1409.3242
  [astro-ph.CO]}}.

\bibitem{Feldman:1993ky}
H.~A. Feldman, N.~Kaiser, and J.~A. Peacock, ``{Power spectrum analysis of
  three-dimensional redshift surveys},''
  \href{http://dx.doi.org/10.1086/174036}{{\em Astrophys. J.} {\bfseries 426}
  (1994) 23--37},
\href{http://arxiv.org/abs/astro-ph/9304022}{{\ttfamily arXiv:astro-ph/9304022
  [astro-ph]}}.

\bibitem{miller74}
L.~D. Miller, M.~E. Miller, and M.~Sivvy, ``The jackknife-a review,'' {\em
  Biometrika} (1974) 1--15.

\bibitem{Norberg2009}
P.~Norberg, C.~M. Baugh, E.~Gaztanaga, and D.~J. Croton, ``{Statistical
  Analysis of Galaxy Surveys - I. Robust error estimation for 2-point
  clustering statistics},''
  \href{http://dx.doi.org/10.1111/j.1365-2966.2009.14389.x}{{\em Mon. Not. Roy.
  Astron. Soc.} {\bfseries 396} (2009) 19},
\href{http://arxiv.org/abs/0810.1885}{{\ttfamily arXiv:0810.1885 [astro-ph]}}.

\bibitem{Norberg:2011ef}
P.~Norberg, E.~Gaztanaga, C.~M. Baugh, and D.~J. Croton, ``{Statistical
  Analysis of Galaxy Surveys-IV: An objective way to quantify the impact of
  superstructures on galaxy clustering statistics},''
  \href{http://dx.doi.org/10.1111/j.1365-2966.2011.19636.x}{{\em Mon. Not. Roy.
  Astron. Soc.} {\bfseries 418} (2011) 2435},
\href{http://arxiv.org/abs/1106.5701}{{\ttfamily arXiv:1106.5701
  [astro-ph.CO]}}.

\bibitem{Guo:2012nk}
H.~Guo {\em et~al.}, ``{The clustering of galaxies in the SDSS-III Baryon
  Oscillation Spectroscopic Survey: Luminosity and Color Dependence and
  Redshift Evolution},''
  \href{http://dx.doi.org/10.1088/0004-637X/767/2/122}{{\em Astrophys. J.}
  {\bfseries 767} (2013) 122},
\href{http://arxiv.org/abs/1212.1211}{{\ttfamily arXiv:1212.1211
  [astro-ph.CO]}}.

\bibitem{Favole:2015rza}
G.~Favole, C.~K. McBride, D.~J. Eisenstein, F.~Prada, M.~E. Swanson, C.-H.
  Chuang, and D.~P. Schneider, ``{Building a better understanding of the
  massive high-redshift BOSS CMASS galaxies as tools for cosmology},''
  \href{http://dx.doi.org/10.1093/mnras/stw1801}{{\em Mon. Not. Roy. Astron.
  Soc.} {\bfseries 462} no.~2, (2016) 2218--2236},
\href{http://arxiv.org/abs/1506.02044}{{\ttfamily arXiv:1506.02044
  [astro-ph.CO]}}.

\bibitem{1982jbor.book.....E}
B.~{Efron}, {\em {The Jackknife, the Bootstrap and other resampling plans}}.
\newblock 1982.

\bibitem{Paz:2015kwa}
D.~J. Paz and A.~G. Sanchez, ``{Improving the precision matrix for precision
  cosmology},'' \href{http://dx.doi.org/10.1093/mnras/stv2259}{{\em Mon. Not.
  Roy. Astron. Soc.} {\bfseries 454} no.~4, (2015) 4326--4334},
\href{http://arxiv.org/abs/1508.03162}{{\ttfamily arXiv:1508.03162
  [astro-ph.CO]}}.

\bibitem{Hartlap:2006kj}
J.~Hartlap, P.~Simon, and P.~Schneider, ``{Why your model parameter confidences
  might be too optimistic: Unbiased estimation of the inverse covariance
  matrix},'' \href{http://dx.doi.org/10.1051/0004-6361:20066170}{{\em Astron.
  Astrophys.} (2006) }, \href{http://arxiv.org/abs/astro-ph/0608064}{{\ttfamily
  arXiv:astro-ph/0608064 [astro-ph]}}.
[Astron. Astrophys.464,399(2007)].

\bibitem{Kaufman:2008}
C.~G. Kaufman, M.~J. Schervish, and D.~W. Nychka, ``Covariance tapering for
  likelihood-based estimation in large spatial data sets,''
  \href{http://dx.doi.org/10.1198/016214508000000959}{{\em Journal of the
  American Statistical Association} {\bfseries 103} no.~484, (2008)
  1545--1555}.

\bibitem{Wendland1995}
H.~Wendland, ``Piecewise polynomial, positive definite and compactly supported
  radial functions of minimal degree,''
  \href{http://dx.doi.org/10.1007/BF02123482}{{\em Advances in Computational
  Mathematics} {\bfseries 4} no.~1, (Dec, 1995) 389--396}.

\bibitem{WENDLAND1998258}
H.~Wendland, ``Error estimates for interpolation by compactly supported radial
  basis functions of minimal degree,''
  \href{http://dx.doi.org/https://doi.org/10.1006/jath.1997.3137}{{\em Journal
  of Approximation Theory} {\bfseries 93} no.~2, (1998) 258 -- 272}.

\bibitem{Padmanabhan:2008ag}
N.~Padmanabhan and M.~J. White, ``{Constraining Anisotropic Baryon
  Oscillations},'' \href{http://dx.doi.org/10.1103/PhysRevD.77.123540}{{\em
  Phys. Rev.} {\bfseries D77} (2008) 123540},
\href{http://arxiv.org/abs/0804.0799}{{\ttfamily arXiv:0804.0799 [astro-ph]}}.

\bibitem{Lesgourgues:2011re}
J.~Lesgourgues, ``{The Cosmic Linear Anisotropy Solving System (CLASS) I:
  Overview},''
\href{http://arxiv.org/abs/1104.2932}{{\ttfamily arXiv:1104.2932
  [astro-ph.IM]}}.

\bibitem{Eisenstein:1997ik}
D.~J. Eisenstein and W.~Hu, ``{Baryonic features in the matter transfer
  function},'' \href{http://dx.doi.org/10.1086/305424}{{\em Astrophys. J.}
  {\bfseries 496} (1998) 605},
\href{http://arxiv.org/abs/astro-ph/9709112}{{\ttfamily arXiv:astro-ph/9709112
  [astro-ph]}}.

\bibitem{Ade:2015xua}
{\bfseries Planck} Collaboration, P.~A.~R. Ade {\em et~al.}, ``{Planck 2015
  results. XIII. Cosmological parameters},''
  \href{http://dx.doi.org/10.1051/0004-6361/201525830}{{\em Astron. Astrophys.}
  {\bfseries 594} (2016) A13},
\href{http://arxiv.org/abs/1502.01589}{{\ttfamily arXiv:1502.01589
  [astro-ph.CO]}}.

\bibitem{Crocce:2005xy}
M.~Crocce and R.~Scoccimarro, ``{Renormalized cosmological perturbation
  theory},'' \href{http://dx.doi.org/10.1103/PhysRevD.73.063519}{{\em Phys.
  Rev.} {\bfseries D73} (2006) 063519},
\href{http://arxiv.org/abs/astro-ph/0509418}{{\ttfamily arXiv:astro-ph/0509418
  [astro-ph]}}.

\bibitem{Xu:2012hg}
X.~Xu, N.~Padmanabhan, D.~J. Eisenstein, K.~T. Mehta, and A.~J. Cuesta, ``{A
  2\% Distance to z=0.35 by Reconstructing Baryon Acoustic Oscillations - II:
  Fitting Techniques},''
  \href{http://dx.doi.org/10.1111/j.1365-2966.2012.21573.x}{{\em Mon. Not. Roy.
  Astron. Soc.} {\bfseries 427} (2012) 2146},
\href{http://arxiv.org/abs/1202.0091}{{\ttfamily arXiv:1202.0091
  [astro-ph.CO]}}.

\bibitem{Hu:1995en}
W.~Hu and N.~Sugiyama, ``{Small scale cosmological perturbations: An Analytic
  approach},'' \href{http://dx.doi.org/10.1086/177989}{{\em Astrophys. J.}
  {\bfseries 471} (1996) 542--570},
\href{http://arxiv.org/abs/astro-ph/9510117}{{\ttfamily arXiv:astro-ph/9510117
  [astro-ph]}}.

\bibitem{Beutler:2011hx}
F.~Beutler, C.~Blake, M.~Colless, D.~H. Jones, L.~Staveley-Smith, L.~Campbell,
  Q.~Parker, W.~Saunders, and F.~Watson, ``{The 6dF Galaxy Survey: Baryon
  Acoustic Oscillations and the Local Hubble Constant},''
  \href{http://dx.doi.org/10.1111/j.1365-2966.2011.19250.x}{{\em Mon. Not. Roy.
  Astron. Soc.} {\bfseries 416} (2011) 3017--3032},
\href{http://arxiv.org/abs/1106.3366}{{\ttfamily arXiv:1106.3366
  [astro-ph.CO]}}.

\bibitem{Abbott:2017wcz}
{\bfseries DES} Collaboration, T.~M.~C. Abbott {\em et~al.}, ``{Dark Energy
  Survey Year 1 Results: Measurement of the Baryon Acoustic Oscillation scale
  in the distribution of galaxies to redshift 1},''
  \href{http://dx.doi.org/10.1093/mnras/sty3351}{{\em Mon. Not. Roy. Astron.
  Soc.} {\bfseries 483} no.~4, (2019) 4866--4883},
\href{http://arxiv.org/abs/1712.06209}{{\ttfamily arXiv:1712.06209
  [astro-ph.CO]}}.

\bibitem{Gil-Marin:2015nqa}
H.~Gil-Marín {\em et~al.}, ``{The clustering of galaxies in the SDSS-III
  Baryon Oscillation Spectroscopic Survey: BAO measurement from the
  LOS-dependent power spectrum of DR12 BOSS galaxies},''
  \href{http://dx.doi.org/10.1093/mnras/stw1264}{{\em Mon. Not. Roy. Astron.
  Soc.} {\bfseries 460} no.~4, (2016) 4210--4219},
\href{http://arxiv.org/abs/1509.06373}{{\ttfamily arXiv:1509.06373
  [astro-ph.CO]}}.

\bibitem{Delubac:2014aqe}
{\bfseries BOSS} Collaboration, T.~Delubac {\em et~al.}, ``{Baryon acoustic
  oscillations in the Ly$\alpha$ forest of BOSS DR11 quasars},''
  \href{http://dx.doi.org/10.1051/0004-6361/201423969}{{\em Astron. Astrophys.}
  {\bfseries 574} (2015) A59},
\href{http://arxiv.org/abs/1404.1801}{{\ttfamily arXiv:1404.1801
  [astro-ph.CO]}}.

\bibitem{chuang2012}
C.-H. Chuang and Y.~Wang, ``{Measurements of H(z) and DA(z) from the
  two-dimensional two-point correlation function of Sloan Digital Sky Survey
  luminous red galaxies},''
  \href{http://dx.doi.org/10.1111/j.1365-2966.2012.21565.x}{{\em Monthly
  Notices of the Royal Astronomical Society} {\bfseries 426} no.~1, (10, 2012)
  226--236}.

\bibitem{Philcox:2019ued}
O.~H.~E. Philcox, D.~J. Eisenstein, R.~O'Connell, and A.~Wiegand, ``{RascalC: A
  Jackknife Approach to Estimating Single and Multi-Tracer Galaxy Covariance
  Matrices},''
\href{http://arxiv.org/abs/1904.11070}{{\ttfamily arXiv:1904.11070
  [astro-ph.CO]}}.

\bibitem{Eisenstein:2006nk}
D.~J. Eisenstein, H.-j. Seo, E.~Sirko, and D.~Spergel, ``{Improving
  Cosmological Distance Measurements by Reconstruction of the Baryon Acoustic
  Peak},'' \href{http://dx.doi.org/10.1086/518712}{{\em Astrophys. J.}
  {\bfseries 664} (2007) 675--679},
\href{http://arxiv.org/abs/astro-ph/0604362}{{\ttfamily arXiv:astro-ph/0604362
  [astro-ph]}}.

\bibitem{Padmanabhan:2012hf}
N.~Padmanabhan, X.~Xu, D.~J. Eisenstein, R.~Scalzo, A.~J. Cuesta, K.~T. Mehta,
  and E.~Kazin, ``{A 2 per cent distance to $z$=0.35 by reconstructing baryon
  acoustic oscillations - I. Methods and application to the Sloan Digital Sky
  Survey},'' \href{http://dx.doi.org/10.1111/j.1365-2966.2012.21888.x}{{\em
  Mon. Not. Roy. Astron. Soc.} {\bfseries 427} no.~3, (2012) 2132--2145},
\href{http://arxiv.org/abs/1202.0090}{{\ttfamily arXiv:1202.0090
  [astro-ph.CO]}}.

\end{thebibliography}\endgroup

\end{document}